\documentclass{iopart}

\usepackage{comment} 
\usepackage{graphicx}
\usepackage[latin1]{inputenc}
\usepackage{color}
\usepackage{iopams}
\usepackage{soul}
\usepackage{cite}

\newcommand{\nc}{\newcommand}
\nc{\text}[1]{{\rm #1}}
\nc{\bra}[1]{\langle #1|}
\nc{\ket}[1]{|#1\rangle}
\nc{\braket}[1]{\left\langle #1 \right\rangle}
\nc{\dagg}{^{\dagger}}
\nc{\conj}{^{*}}
\nc{\dx}[1]{\, \mathrm{d} {#1} \,}
\nc{\Dx}[1]{\mathcal{D} {#1} \,}
\nc{\la}{\langle}
\nc{\ra}{\rangle}
\nc{\Id}{\mathbb{1}}
\nc{\eps}{\varepsilon}
\nc{\der}[2]{\frac{\mathrm{d} {#1}}{\mathrm{d} {#2}}}
\nc{\pder}[2]{\frac{\partial {#1}}{\partial {#2}}}
\nc{\bigO}{\mathcal{O}}
\nc{\Eq}[1]{Equation (\ref{#1})}
\nc{\eq}[1]{Eq. (\ref{#1})}
\nc{\chap}[1]{Chapter \ref{#1}}
\nc{\sect}[1]{Section \ref{#1}}
\nc{\fig}[1]{figure \ref{#1}}
\nc{\Fig}[1]{Fig. \ref{#1}}
\nc{\tabl}[1]{Table \ref{#1}}
\nc{\app}[1]{Appendix \ref{#1}}
\renewcommand{\Re}{\text{Re}\,}
\renewcommand{\Im}{\text{Im}\,}
\nc{\half}{\frac{1}{2}}

\nc{\alg}[1]{\textcolor{blue}{#1}}
\nc{\asp}[1]{#1}

\begin{document}

\title[]{Rapid Steady State Convergence for Quantum Systems Using Time-Delayed Feedback Control}

\author{A.L. Grimsmo$^{1}$, A.S. Parkins$^2$ and B.-S. Skagerstam$^{1}$}
\address{$^1$ Department of Physics, The Norwegian University of Science and Technology, N-7491 Trondheim, Norway}
\address{$^2$ Department of Physics, University of Auckland, Private Bag 92019, Auckland, New Zealand}
\ead{arne.grimsmo@ntnu.no}
\ead{s.parkins@auckland.ac.nz}
\ead{bo-sture.skagerstam@ntnu.no}



\date{\today}

\begin{abstract}
We propose a time-delayed feedback control scheme for open quantum systems that can dramatically reduce the time to reach steady state. No measurement is performed in the feedback loop, and we suggest a simple all-optical implementation for a cavity QED system. We demonstrate the potential of the scheme by applying it to a driven and dissipative Dicke model, as recently realized in a quantum gas experiment. The time to reach steady state can then reduced by two orders of magnitude for parameters taken from experiment, making previously inaccessible long time attractors reachable within typical experimental run times. The scheme also offers the possibility of slowing down the dynamics, as well as qualitatively changing the phase diagram of the system.
\end{abstract}


\maketitle


\section{Introduction}

Steady states of open quantum systems, where driving forces and internal dynamics are balanced by dissipation and/or other types of environmental noise, are often of experimental interest. There are indeed a number of platforms currently available, where the experimental control of the individual constituents of interacting quantum systems allows precise preparation of interesting and non-trivial steady states, through measurement and control of well-defined outputs and inputs of the system. These include systems using trapped ions or ultra-cold atoms, opto-mechanical systems, and systems based on cavity or circuit quantum electrodynamics (CQED) (see e.g. Refs.\cite{Muller12,Kippenberg08,Mabuchi02}). This can be an alternative to quantum state preparation by coherent (unitary) evolution, and is potentially of great interest to quantum information processing technologies as a way of preparing computational resources, such as maximally entangled states \cite{Kastoryano11,Krauter11,Shankar13,Lin13}, or even providing for a route to quantum computation \cite{Verstraete09}. One great advantage of such an approach is that the steady state is robust against variations of the initial state. 

In practice, the time-scale on which such a steady state is reached is very important; the generation of the desired state often requires a degree of control that is hard to sustain over time, which can pose a challenge for finite-time experiments \cite{Baumann10,Keeling10,Bhaseen12}. In the present paper, we propose an all-optical feedback scheme relevant, for example, to CQED systems, that can be used to $i)$ change the stability of long time attractors, so that  one can switch between different behaviors, and $ii)$ change the characteristic time-scale for approaching a steady state, thus potentially speeding up the convergence. The scheme we use is based on the time-delayed feedback control method developed by Pyragas \cite{Pyragas92}, and often referred to as the time-delay auto-synchronization (TDAS) \cite{Socolar94,Gauthier94}. The control is based on \emph{coherent} feedback, i.e., no measurement is performed in the feedback loop, which can be advantageous, or even necessary, for stabilizing the high frequency dynamics of optical systems or high speed electrical circuits \cite{Socolar94,Gauthier94}. Another great strength of the approach is that it does not require the steady state to be known a priori. We apply coherent TDAS, to our knowledge, for the first time to a quantum system, with the feedback signal treated quantum mechanically as well \cite{Wiseman94,Lloyd00,Nurdin09}. Delay-times have often been assumed to be negligible in theoretical modelling of coherent quantum feedback, motivated by the fact that a time-delay can introduce undesirable instability to the system, and that it can often be made very small in practice \cite{Wiseman94,Wiseman10}. In contrast, with TDAS, the delay is the crucial ingredient for \emph{increasing} the system's stability. 
We note that delayed coherent feedback has also been considered in \cite{Carmele13,Zhang13} for a single-atom single-excitation system (and under the usual rotating wave approximation).

We will demonstrate the potential of our scheme by applying it to a highly topical example, namely an open system version of the Dicke model. The Dicke model is a paradigmatic model in quantum optics, describing the interaction of a collection of two-level atoms with a single cavity mode \cite{Dicke54}. This model has been realized and studied in a number of recent experiments \cite{Baumann10,Baumann11,Brennecke13} based on a Bose-Einstein condensate (BEC) coupled to a field mode of a high finesse optical cavity. The cavity has a natural dissipative output channel, which has been used to monitor the system in real time. In particular, the system undergoes a quantum phase transition as an effective coupling strength between the BEC and the light field is increased beyond a critical value, which can be observed through the intensity of the output cavity field \cite{Baumann10}. Spontaneous symmetry breaking has also been observed through a heterodyne measurement scheme \cite{Baumann11}, and measuring the correlations of the density fluctuations has been used to observe the diverging time-scale upon approaching the critical point \cite{Brennecke13}. This type of monitoring of a dissipative channel is non-destructive, and as these experiments have shown, offers a very promising route for the observation of complex many-body quantum dynamics. 

We will take advantage of this dissipative channel as well, by using it as the input to a non-invasive feedback loop that can alter the characteristic time-scale for the relaxation of the system, as well as the stability of the long-time attractors. This is particularly relevant for the BEC realization of the Dicke model where, as we will discuss in more detail below, the approach to steady state can be slow compared to typical experimental run times.  Adverse effects such as spontaneous emission, or atom loss, will eventually cause deviations from the desired, idealized behavior. This is particularly problematic close to phase boundaries, where we expect critical slowing down. The system is therefore a very interesting test-case for our proposed feedback scheme. We will treat the feedback in a semi-classical approach, where quantum fluctuations are linearized. For the Dicke model, this approximation is valid in the thermodynamic limit of a large number of atoms. The approach we develop should be similarly applicable to a variety of topical quantum-optical systems for which feedback can also have useful and interesting consequences
 (see e.g. Refs. \cite{Gough09,Iida12,Hamerly12,Mabuchi11}).

The paper is organized as follows.  In \sect{sect:fbcontrol} we introduce our feedback scheme in a general setting, using the standard input-output theory for quantum optical systems. Then, in \sect{sect:mean}, we introduce our primary system of study, the Dicke model as recently experimentally realized, and apply our feedback scheme. We study in detail the effect of the forcing due to the feedback, and find optimal delay-times for rapid convergence to steady state. We compare the performance of the system with and without feedback, and demonstrate improvements in the relaxation time of two orders of magnitude. In \sect{sect:sweep} we consider potential consequences for finite-time experiments. In \sect{sect:quantumfluct} we consider the influence of the feedback force on quantum fluctuations. Finally, in \sect{sect:conc}, we give some concluding remarks.

\section{\label{sect:fbcontrol}Coherent time-delay auto-synchronization}

The Pyragas' time-delay auto-synchronization (TDAS) method is a continuous feedback control method first developed to stabilize unstable periodic orbits and equilibrium states embedded in a chaotic attractor \cite{Pyragas92,Socolar94,Gauthier94,Ahlborn04}. We will  also use the method for manipulating the characteristic time-scale on which a stable fixed point is approached. 
Briefly, the idea behind TDAS for stabilizing a dynamical system, whose classical state is given by $x(t)$, is to apply one or more continuous feedback forces of the form $F(t) = k[x(t)-x(t-\tau)]$. The feedback force vanishes in steady state, or for a periodic orbit if the delay, $\tau$, is a multiple of the period. This is referred to as \emph{non-invasive} feedback. The delay, and feedback strength, $k$, are parameters that should be varied in experiment to achieve driving towards a particular long-time attractor.

%
\begin{figure}
\vspace{10mm}
\begin{picture}(0,0)(-50,150)   
\includegraphics{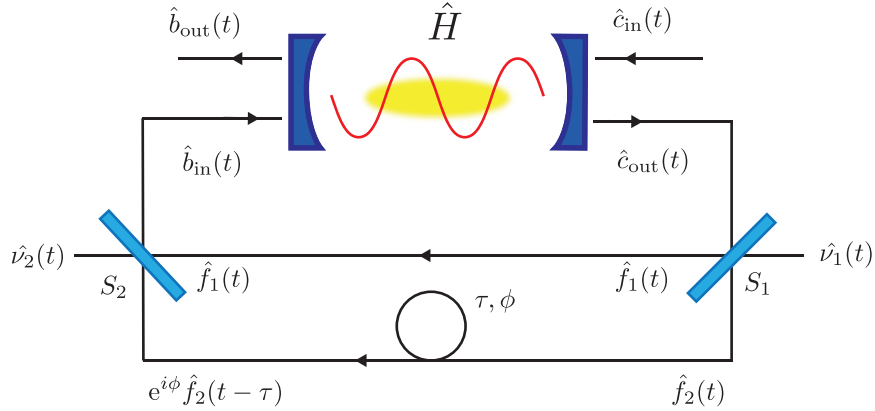}
\end{picture}
\vspace{60mm}
  \caption{\label{fig:feedback}  A schematic illustration of our feedback control scheme. The internal dynamics of the cavity is described by a Hamiltonian, $\hat{H}$. The cavity consists of two mirrors, $b$ (left) and $c$ (right), with decay rates $\kappa_b$ and $\kappa_c$ respectively. The light-blue tilted bars denote beam splitters $BS1$ and $BS2$, with transformation properties defined by the unitary matrices $S_{1}$ and $S_{2}$, respectively. The circle denotes a delay-time of $\tau$ and a $\phi$ phase shift to one of the feedback arms. We assume that Faraday isolators (not shown) separate the cavity input and output fields.}
\end{figure}
%
We consider a CQED system, as illustrated in \Fig{fig:feedback}. The internal dynamics of a cavity, consisting of a field mode, possibly interacting with other quantum and classical degrees of freedom, is described by a Hamiltonian $\hat{H}$. The cavity is assumed to have two mirrors $b$ and $c$, corresponding to two distinct pairs of input-output ports. The equation of motion for the cavity mode is, according to the standard input-output theory for optical quantum systems \cite{Gardiner85},
\begin{eqnarray}\label{eq:inputoutput}
  \frac{d\hat{a}}{dt} = i[\hat{H},\hat{a}] - (\kappa_b + \kappa_c) \hat{a} 
                 - \sqrt{2\kappa_b} \hat{b}_\text{in}(t) - \sqrt{2\kappa_c} \hat{c}_\text{in}(t)\, .
\end{eqnarray}
Here $\hat{a}$ is the cavity mode annihilation operator, $\kappa_{b}$ and $\kappa_{c}$   are the decay rates for the two mirrors, and $\hat{b}_\text{in}(t)$, $\hat{c}_\text{in}(t)$ the annihilation operators of the input fields incident on the respective mirrors. Here, and in the following, we will, when convenient,  suppress the time argument for any system operator evaluated at time $t$, but keep the time argument for input and output fields for clarity. We will assume that $\hat{c}_\text{in}(t)$ corresponds to a vacuum field, although it would also be of interest to consider a drive here. The input fields obey the commutation relations
\begin{eqnarray}\label{eq:input_comm}
  [\hat{\xi}_\text{in}(t),\hat{\xi}\dagg_\text{in}(t')] = \delta(t-t')\, ,
\end{eqnarray}
where $\hat{\xi}_\text{in}$ denotes any of the input mode operators $\hat{b}_\text{in}$ or $\hat{c}_\text{in}$. In addition, any \emph{vacuum} input field $\hat{\xi}_\text{in}(t)$ corresponds to a Gaussian white noise operator with zero mean, and its only non-vanishing correlation function is
\begin{eqnarray}\label{eq:input_vac}
  \braket{\hat{\xi}_\text{in}(t)\hat{\xi}_\text{in}\dagg(t')} &= \delta(t-t')\, .
\end{eqnarray}
The corresponding output fields are given as
\numparts
\begin{eqnarray}
  \hat{b}_\text{out}(t) = \sqrt{2\kappa_b} \hat{a}(t) + \hat{b}_\text{in}(t)\, ,
\end{eqnarray}
and
\begin{eqnarray}
  \hat{c}_\text{out}(t) = \sqrt{2\kappa_c} \hat{a}(t) + \hat{c}_\text{in}(t)\, .
\end{eqnarray}
\endnumparts
The output from mirror $c$ is now split into two feedback arms, as illustrated in \Fig{fig:feedback}. One of the feedback arms has a time-delay $\tau$ and a phase shift $\phi$, while we set the time-delay and phase shift of the other arm to zero for simplicity.  The input and output fields for the two beam-splitters $BS1$ and $BS2$ shown in \Fig{fig:feedback} are related through
\numparts
\begin{eqnarray}
  \left(\begin{array}{c}
      \hat{f}_1(t) \\
      \hat{f}_2(t)
  \end{array}\right)
  = S_1
  \left(\begin{array}{c}
      \hat{\nu}_1(t) \\
      \hat{c}_\text{out}(t)
  \end{array}\right)\, ,
\end{eqnarray}
and
\begin{eqnarray}
  \left(\begin{array}{c}
      \hat{b}_\text{in}(t) \\
      \hat{\nu}_2(t)
  \end{array}\right)
  = S_2
  \left(\begin{array}{c}
      \e^{i\phi}\hat{f}_2(t-\tau) \\
      \hat{f}_1(t)
  \end{array}\right)\, ,
\end{eqnarray}
\endnumparts
where $S_{1}$ and $S_{2}$  are unitary matrices, $\hat{\nu}_1(t)$ is a vacuum input field to beam splitter $BS1$, and $\hat{\nu}_2(t)$ is the (unused) other output field from beam splitter $BS2$.
We want the time-delayed and time-undelayed fields to be incident on mirror $b$  in opposite phase, so as to give the desired destructive interference in steady state. For this purpose, we choose the beam splitter transformations to be given by
\numparts
\begin{eqnarray}
  S_1 =& \frac{\e^{-i\phi/2}}{\sqrt{2}}\left(\begin{array}{cc}
  s & -r\e^{i\phi/2} \\
  r\e^{-i\phi/2} & s
  \end{array}\right)\, , 
\end{eqnarray}
and
\begin{eqnarray}
  S_2 =& \frac{\e^{-i\phi/2}}{\sqrt{2}}\left(\begin{array}{cc}
  -r & -s\e^{i\phi/2} \\
  s\e^{-i\phi/2} & -r
  \end{array}\right)\, ,
\end{eqnarray}
\endnumparts
where $r,s \ge 0$ (real) and $r^2 + s^2 = 2$. In passing, we remark that these are not the most general choice of beam splitter transformations but sufficient for our purposes. We now find for $\hat{b}_\text{in}(t)$:
\begin{eqnarray}\label{eq:bin}
\hspace{28mm} \hat{b}_\text{in}(t) =\nonumber \\ \frac{rs}{2}\left[\hat{c}_\text{out}(t) - \hat{c}_\text{out}(t-\tau)\right] - \frac{s^2}{2}\e^{-i\phi/2} \hat{\nu}_1(t) - \frac{r^2}{2}\e^{-i\phi/2} \hat{\nu}_1(t-\tau) \nonumber \\
                       = \frac{rs}{2}\sqrt{2\kappa_c}\left[\hat{a}(t) - \hat{a}(t-\tau)\right] + \tilde{b}_\text{in}(t),
\end{eqnarray}
where we have defined
\begin{eqnarray}\label{eq:tildebin}
 ~~~~~~~~~~~~~~~~~~~~ \tilde{b}_\text{in}(t) \equiv  \nonumber \\
\frac{rs}{2}\left[\hat{c}_\text{in}(t)- \hat{c}_\text{in}(t-\tau)\right] 
                         - \frac{s^2}{2}\e^{-i\phi/2}\hat{\nu}_1(t) - \frac{r^2}{2}\e^{-i\phi/2}\hat{\nu}_1(t-\tau)\, . 
\end{eqnarray}
The mode operator $\tilde{b}_\text{in}(t)$ satisfies \eq{eq:input_comm} and \eq{eq:input_vac} and is thus the ``vacuum part'' of the field incident on mirror $b$.

A control force is generated from the difference between the current cavity field, ${\hat a}(t)$, and the field at some point in the past, ${\hat a}(t-\tau)$. This forcing is then fed back into the system. By making use of  \eq{eq:bin} in \eq{eq:inputoutput} we obtain
\begin{eqnarray}\label{eq:fb}
  \frac{d\hat{a}(t)}{dt} &=& 
i[\hat{H},\hat{a}(t)] - (\kappa_b + \kappa_c) \hat{a}(t)
                  -\sqrt{2\kappa_b} \tilde{b}_\text{in}(t)\nonumber - \sqrt{2\kappa_c} \hat{c}_\text{in}(t)\\ 
                 &&+ k \left(\hat{a}(t-\tau)-\hat{a}(t)\right) \, ,
\end{eqnarray}
where $k \equiv rs\sqrt{\kappa_b \kappa_c}$ satisfies $0\le k \le \sqrt{\kappa_b\kappa_c}$. We remark that the vacuum input fields $\tilde{b}_\text{in}(t)$ and $\hat{c}_\text{in}(t)$ are not independent, as can be seen from \eq{eq:tildebin}.

In the following section we will apply this scheme to an open version of the Dicke model, and find optimal delay-times for parameters motivated by recent experiment.

\section{\label{sect:mean}Application to the open Dicke model dynamics}

We will first introduce the generalized Dicke model that will be the main object of our study. It describes the interaction of $N$ two-level atoms with a single mode of the electro-magnetic field.  The Hamiltonian describing the internal dynamics of the system is
\begin{eqnarray}\label{eq:H_Dicke}
  \hat{H} =& \omega_0 \hat{J}_z + \omega \hat{a}\dagg \hat{a} + \frac{2g}{\sqrt{N}} \hat{J}_x \left(\hat{a} + \hat{a}\dagg\right) + \frac{U}{N} \hat{J}_z \hat{a}\dagg \hat{a}\, ,
\end{eqnarray}
where parameters $\omega_0$ and $\omega$ are atomic and cavity frequencies, respectively, $g$ is the linear interaction strength, and $U$ is a non-linear coupling constant. The operators $\hat{a}$ and $\hat{a}\dagg$ are, again, the annihilation and creation operators for the cavity mode, and $\{\hat{J}_x, \hat{J}_y, \hat{J}_z\}$ are collective atomic operators satisfying the conventional angular momentum commutation relations. The Hamiltonian in \eq{eq:H_Dicke} is identical to the conventional Dicke model \cite{Dicke54} Hamiltonian  if $U$ is set to zero.

We will not enter into the details on the underlying physics of the BEC experiments realizing \eq{eq:H_Dicke}, but refer the reader to Ref. \cite{Baumann10}. Briefly, the two-level atoms of the Dicke model are realized through pairs of discrete momentum states of a BEC. The linear coupling to the cavity field, $(2g/\sqrt{N}) \hat{J}_x (\hat{a}\dagg + \hat{a})$, effectively describes Rayleigh scattering of photons between the cavity mode and an auxiliary pump laser.
Importantly, the effective coupling strength, $g \sim \sqrt{P}$, can be tuned via the laser intensity $P$. The parameter $\omega$ is determined by the detuning of the cavity mode frequency from the pump laser frequency, and is therefore  controllable. The collective atomic frequency $\omega_0$ is fixed by the optical wave-vector and atomic mass (i.e., it is set by the recoil energy). It is therefore not readily tunable like the other parameters. The non-linear coupling constant $U$ is given by a dispersive 
light-shift. We will consider parameters where this non-linear coupling does not play a major role, but refer to 
Refs.\cite{Keeling10,Bhaseen12,Grimsmo13} for discussions on the very interesting dynamics that can result from this term.

We will assume that the cavity consists of two mirrors, both with decay rates $\kappa_b=\kappa_c\equiv\kappa/2$ (for simplicity), and implement the feedback scheme introduced in the previous section and illustrated in \Fig{fig:feedback}. We will, furthermore, assume the ``thermodynamic'' limit $N \to \infty$, where we can employ a semi-classical approach to find expectation values of system operators, as described below. Later in \sect{sect:quantumfluct}, we examine quantum fluctuations by linearizing the operator equations of motion around these expectation values.

By making use of \eq{eq:fb} and \eq{eq:H_Dicke}, we can now write down the Heisenberg equations of motion for the cavity mode and the spin operators:
\begin{eqnarray}\label{eq:inoutdicke}
  \frac{d\hat{a}(t)}{dt} =& -i\left(\omega + \frac{U}{N} \hat{J}_z(t)\right)\hat{a}(t) - i\frac{2g}{\sqrt{N}} \hat{J}_x(t) - \kappa \hat{a}(t) \nonumber \\
                 &+ k\left(\hat{a}(t-\tau) - \hat{a}(t)\right) -\sqrt{2\kappa} \hat{a}_\text{in}(t) \nonumber\, , \\
  \frac{d\hat{J}_x(t)}{dt} =& -\left(\omega_0 + \frac{U}{N} \hat{a}(t)\dagg \hat{a}(t)\right) \hat{J}_y(t)\, , \nonumber\\
  \frac{d\hat{J}_y(t)}{dt}=& \left(\omega_0 + \frac{U}{N} \hat{a}(t)\dagg \hat{a}(t)\right) \hat{J}_x(t) - \frac{2g}{\sqrt{N}}\left(\hat{a}(t) + \hat{a}(t)\dagg\right) \hat{J}_z(t) \nonumber\, , \\
  \frac{d\hat{J}_z(t)}{dt} =& \frac{2g}{\sqrt{N}}\left(\hat{a}(t) + \hat{a}(t)\dagg\right) \hat{J}_y(t) \, ,
\end{eqnarray}
where now $0 \le k \le \kappa/2$, and $\hat{a}_\text{in}(t) = 1/\sqrt{2}\left(\tilde{b}_\text{in}(t)+\hat{c}_\text{in}(t)\right)$ is used to denote the sum of the vacuum input fields through the two mirrors and where, for clarity, the time-dependence has been made explicit. Since $\tilde{b}_\text{in}(t)$, which corresponds to a vacuum part of the field incident on mirror $b$, is not independent of the input field $\hat{c}_\text{in}(t)$ on mirror $c$, $\hat{a}_\text{in}(t)$ does not obey the usual relations \eq{eq:input_comm} and \eq{eq:input_vac}, but instead we have that
\begin{eqnarray}\label{eq:corr_ain}
  [\hat{a}_\text{in}(t),\hat{a}_\text{in}\dagg(t')] = \braket{\hat{a}_\text{in}(t)\hat{a}_\text{in}\dagg(t')}   =\delta(t-t') \nonumber \\
 +
 \frac{k}{\kappa}\left[\delta(t-t')-\half\delta(t-t'+\tau)-\half\delta(t-t'-\tau)\right]\, ,
\end{eqnarray}
which can be verified by making use of  \eq{eq:tildebin}.

The non-linear operator equations, \eq{eq:inoutdicke}, can not be solved directly, and with delayed feedback ($\tau\ne 0$), no equivalent master equation can be derived. By neglecting fluctuations and factorizing operator products, we can, however, derive a closed set of equations of motion for the five real-valued variables:
\begin{eqnarray}
  x_1 \equiv \Re \frac{\braket{\hat{a}}}{\sqrt{N}}\, ,\qquad x_2 \equiv \Im \frac{\braket{\hat{a}}}{\sqrt{N}}\, ,\nonumber \\
  j_x \equiv \frac{\langle\hat{J}_x\rangle}{N}\, ,\qquad j_y \equiv \frac{\langle\hat{J}_y\rangle}{N}\, , \qquad j_z \equiv \frac{\langle\hat{J}_z\rangle}{N}\, , 
\end{eqnarray}
which take the form
\begin{eqnarray}\label{eq:semiclass}
  \frac{dx_1(t)}{dt} =& -\kappa x_1(t) + \big(\omega + Uj_z(t)\big)x_2(t)
             + k\big((x_1(t-\tau) - x_1(t)\big)\nonumber,\\
  \frac{dx_2(t)}{dt} =& -\kappa x_2(t) - \big(\omega + Uj_z(t)\big)x_1(t) - 2gj_x(t)
             + k\big(x_2(t-\tau) - x_2(t)\big)\, , \nonumber\\
  \frac{dj_x(t)}{dt} =& -\big(\omega_0 + U(x_1^2(t) + x_2^2(t))\big) j_y\, ,\nonumber\\
  \frac{dj_y(t)}{dt}  =& \big(\omega_0 + U\big(x_1^2(t) + x_2^2(t)\big)\big) j_x(t) - 4gx_1(t)j_z(t)\, ,\nonumber\\
  \frac{dj_z(t)}{dt}  =& 4gx_1(t)j_y(t)\, .
\end{eqnarray}
Here, again, the time-dependence has been made explicit for clarity. These equations of motion conserve the total length of the spin $j_x^2 + j_y^2 + j_z^2 $, and we will restrict our study to states on the Bloch sphere and thus always assume the constraint
\begin{eqnarray}\label{eq:constraint}
  j_x^2 + j_y^2 + j_z^2 = 1/4\, .
\end{eqnarray}
Below we will also  make use the rescaled complex variable $\alpha \equiv \braket{\hat{a}}/\sqrt{N} = x_1 + ix_2$, and, for notational convenience, we  also write $\mathbf{x} \equiv (x_1,x_2,j_x,j_y,j_z)$.

In the absence of feedback, i.e. $k=0$, this model has been explored theoretically in great detail, both  in the thermodynamic limit $N\to\infty$ in Refs.\cite{Keeling10,Bhaseen12}, and for finite $N$ and including all quantum effects in Ref.\cite{Grimsmo13}. In Refs.\cite{Keeling10,Bhaseen12}, steady states were found analytically by setting the left-hand sides in \eq{eq:semiclass} equal to zero, and a further stability analysis was performed by linearizing around the fixed points.  A surprisingly rich phase diagram was uncovered, with the appearance of several new phases due to the presence of the non-linear coupling $U$ and the cavity decay parameter $\kappa$ when compared to the conventional Dicke model \cite{Emary03a,Emary03b}. One of the key findings in Ref.\cite{Bhaseen12} was that the emergent time-scales of the collective dynamics vary significantly throughout the phase diagram, and that the run-times of current experiments may not be sufficient to reach the long time attractor in all cases.

\begin{figure}
\vspace{50mm}
\begin{picture}(0,0)(-50,0)   
\includegraphics{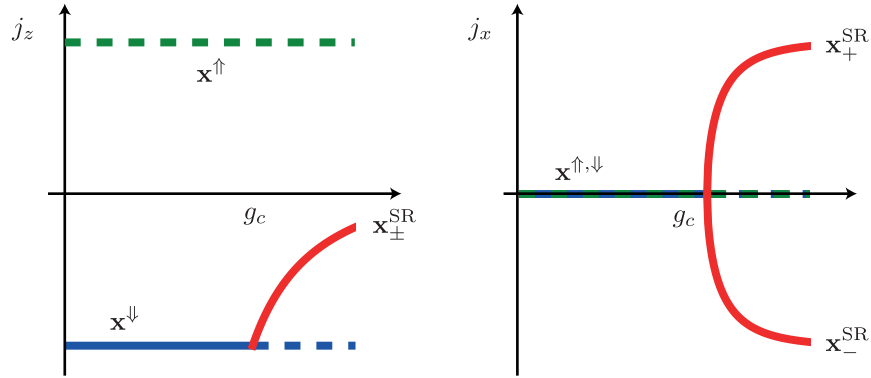}
\end{picture}
  \caption{\label{fig:order} Qualitative illustration of the order parameters $j_z$ ({\bf left figure}) and $j_x$ ({\bf right figure}), as functions of $g$, in the absence of feedback ($k=0$). Below critical coupling, $g<g_c$, there are two fixed points $\mathbf{x}^\Uparrow$ (green) and $\mathbf{x}^\Downarrow$ (blue), where only the latter is stable. Above critical coupling, $g>g_c$, they are both unstable, and two new stable fixed points, $\mathbf{x}^\text{SR}_\pm$ (red), come into existence, satisfying \eq{eq:SR}. Note that the positions of fixed points stay the same under TDAS feedback control, only their stability might change.
  }
\end{figure}

Here we will now focus our attention on the following set of parameters as taken from the recent experiment in Ref.\cite{Brennecke13}:
\begin{eqnarray}\label{eq:params}
  \{\omega_0,\omega,U,\kappa\} = \{ 8.3 \cdot 10^{-3}, 14.0, -8.0, 1.25\} \cdot 2\pi\, \text{ MHz}\, .
\end{eqnarray}
With this choice of parameters, a phase transition from the normal phase, $\mathbf{x}^\Downarrow \equiv (0,0,0,0,-1/2)$, to a super-radiant phase with $\{x_1,x_2,j_x\} \ne 0$, happens at a critical coupling strength \cite{Bhaseen12}
\begin{eqnarray}
  g_c = \sqrt{\frac{\omega_0[(\omega - U/2)^2+\kappa^2]}{4(\omega - U/2)}} = 0.19 \cdot 2\pi\, \text{ MHz}\, .
\end{eqnarray}
Below this critical coupling there are two fixed points: the normal phase
\begin{eqnarray}\label{eq:norm}
  \mathbf{x}^\Downarrow \equiv (0,0,0,0,-1/2)\, ,
\end{eqnarray}
and the inverted phase
\begin{eqnarray}\label{eq:inv}
  \mathbf{x}^\Uparrow\equiv(0,0,0,0,1/2)\, , 
\end{eqnarray}
where only the former is stable in the absence of feedback ($k=0$). Above the critical coupling, both of these phases are unstable in the absence of feedback, and two new stable fixed points, $\mathbf{x}^\text{SR}_\pm$, come into existence, given by \cite{Bhaseen12}
\begin{eqnarray}\label{eq:SR}
  &j_z^{\text{SR}} = \left\{\begin{array}{ll}
  -\displaystyle{\frac{\omega}{U}} - \sqrt{\frac{g^2(4\omega^2-U^2)-\omega_0 U \kappa^2}{U^2(\omega_0 U + 4g^2)}}\, , &\text{if }\, U \ne 0\, , \\
  - \displaystyle{\frac{g_c^2}{2g^2}}\, , &\text{if }\, U = 0\, ,
\end{array}\right.\nonumber \\
&j_x^{\text{SR}} = \pm \sqrt{1/4-(j_z^\text{SR})^2}\, ,\nonumber \\
  &j_y^{\text{SR}} = 0\, , \nonumber\\
  &\alpha^{\text{SR}} \equiv x_1^{\text{SR}} + i x_2^{\text{SR}} = -\frac{2gj_x^{\text{SR}}}{\omega+U j_z^{\text{SR}} -i\kappa}\, .
\end{eqnarray}
The two fixed points $\mathbf{x}^\text{SR}_\pm$ differ only in the choice of sign for $j_x$ and $\alpha$, related to a duality emerging from the invariance of \eq{eq:H_Dicke} under the parity transformation $\hat{a} \to -\hat{a}, \hat{J}_x \to -\hat{J_x}$. We can treat the two solutions simultaneously, and we will refer to both as ``the super-radiant phase'', and denote them both by $\mathbf{x}^\text{SR}$, when the difference between them is of no importance. The super-radiant phase transition (in the absence of feedback) is qualitatively illustrated in \Fig{fig:order}.

\begin{figure}
\vspace{100mm}
\begin{picture}(0,0)(-20,0)   
\includegraphics{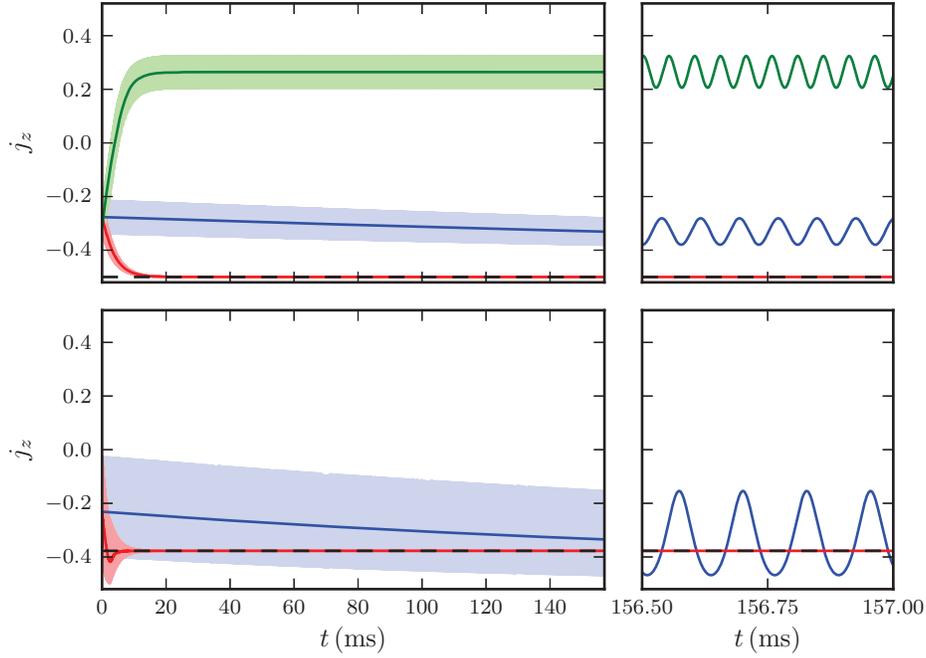}
\end{picture}
  \caption{\label{fig:timeevol} Time-evolution of the collective inversion, $j_z(t)$, for an initial state $\mathbf{x}=(0,0,1/\sqrt{12},1/\sqrt{12},1/\sqrt{12})$. The Hamiltonian coupling of the spin and cavity degrees of freedom induces oscillations that are very rapid compared to the relaxation time. The shaded regions in the left panels show the oscillating solution, while the lines show the same signal after a low-pass filter has been applied. The right panels show a zoom of the oscillating solutions for late times. The dashed lines show the exact steady state values for $k=0$. {\bf Top panels}: $g/g_c = 0.74$; the normal phase, $\mathbf{x}^\Downarrow$ as in \eq{eq:norm}, is the only stable fixed point in the absence of feedback. The blue line shows the time-evolution without feedback. The red line is with $k=\kappa/2$ and $\tau=50\,\mu$s. The green line shows the time-evolution for $k=\kappa/2$ and $\tau=100\,\mu$s, for which the normal phase is unstable, and the system exhibits persistent oscillations in the long time limit. {\bf Bottom panels}: $g/g_c = 1.1$; the super-radiant phase, $\mathbf{x}^\text{SR}$ as in \eq{eq:SR}, is the only stable fixed point in the absence of feedback. The blue line shows the time-evolution without feedback, and the red line is for $k=\kappa/2$ and $\tau = 50\,\mu$s. For both panels, all other parameters are given in \eq{eq:params}.}
\end{figure}

\begin{figure}
\vspace{100mm}
\begin{picture}(0,0)(-20,0)   
\includegraphics{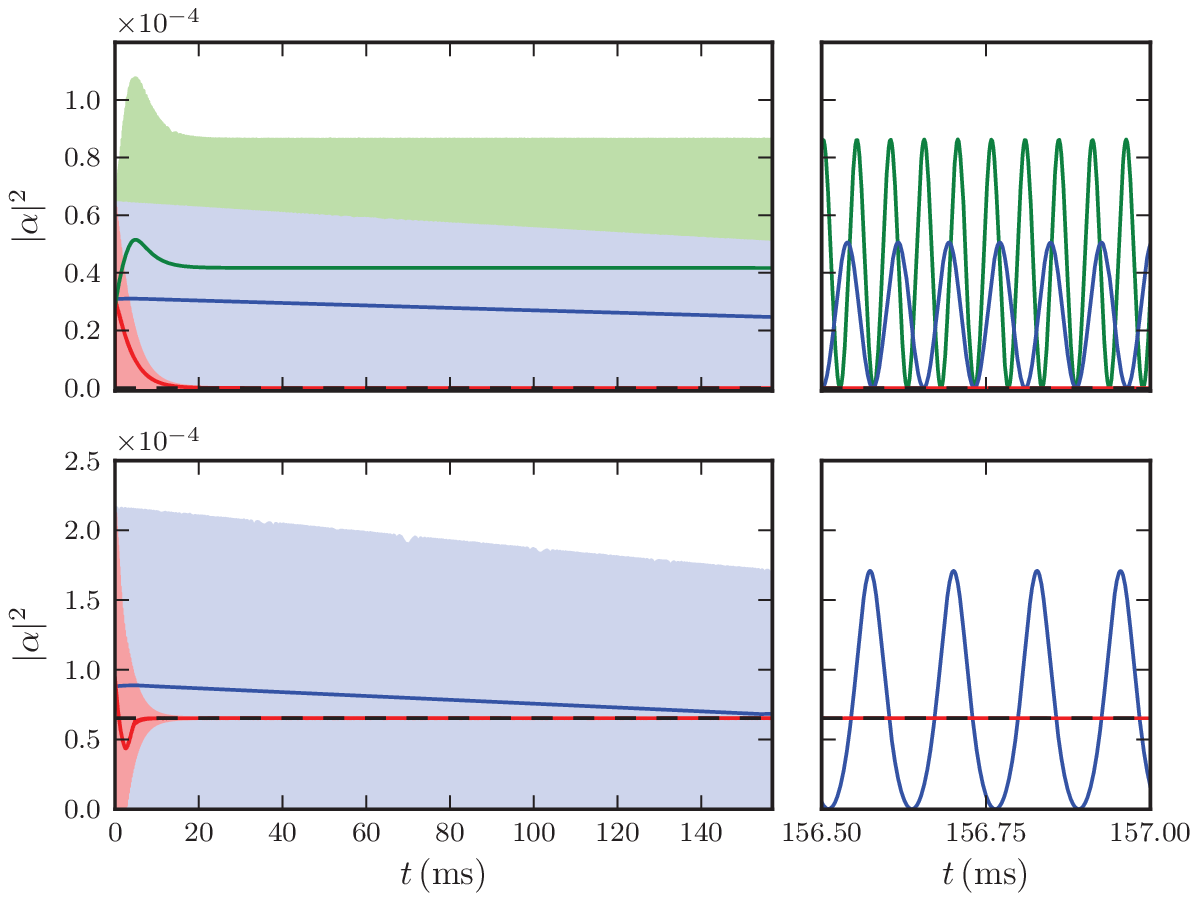}
\end{picture}
  \caption{\label{fig:timeevol_ada} Time-evolution of the (normalized) photon number, $|\alpha(t)|^2 \equiv |x_1(t)|^2 + |x_2(t)|^2$, for an initial state $\mathbf{x}=(0,0,1/\sqrt{12},1/\sqrt{12},1/\sqrt{12})$. The shaded regions in the left panels show the oscillating solution, while the lines show the same signal after a low-pass filter has been applied. The right panels show a zoom of the oscillating solutions for late times. The dashed lines show the exact steady state values for $k=0$. {\bf Top panels}: $g/g_c = 0.74$; the normal phase, $\mathbf{x}^\Downarrow$ \eq{eq:norm}, is the only stable fixed point in the absence of feedback. The blue line shows the time-evolution without feedback. The red line is with $k=\kappa/2$ and $\tau=50\,\mu$s. The green line shows the time-evolution for $k=\kappa/2$ and $\tau=100\,\mu$s, for which the normal phase is unstable, and the system exhibits persistent oscillations in the long time limit. {\bf Bottom panels}: $g/g_c = 1.1$; the super-radiant phase, $\mathbf{x}^\text{SR}$ \eq{eq:SR}, is the only stable fixed point in the absence of feedback. The blue line shows the time-evolution without feedback, and the red line is for $k=\kappa/2$ and $\tau = 50\,\mu$s. For both panels, all other parameters are given in \eq{eq:params}.}
\end{figure}

The treatment in Refs.\cite{Keeling10,Bhaseen12} showed that the approach to steady state can be exceedingly slow. This is partly related to the relatively small value of $\omega_0$, and can in fact be interpreted as critical slowing down, due to closeness to phase boundaries in an extended parameter space \cite{Keeling10,Bhaseen12}. By careful adiabatic elimination of the cavity mode, under the condition $\{\omega,\kappa\} \gg \omega_0$, one can find an effective rate describing the incoherent dynamics of the atoms. In the normal phase, for example, it is found  to be $\simeq 4\kappa g^2 (\omega-U/2)\omega_0/\left[(\omega-U/2)^2+\kappa^2\right]^2$ \cite{Bhaseen12} (see also equation \eq{eq:lambda_approx_b} below). At $g=g_c$ this is roughly $0.3 \cdot 2\pi$ Hz, which indicates a remarkably slow decay taking on the order of seconds. The parameter $\omega_0$, describing the collective atomic frequency, is fixed by the optical wave vector and the atomic mass, and can therefore not easily be made larger in practice. In \Fig{fig:timeevol}, we show two typical examples of the atomic inversion, $j_z(t)$, as a function of time, below critical coupling, with $g/g_c = 0.74$ and above, with $g/g_c = 1.1$. The other parameters are as in \eq{eq:params}. In both cases the initial state is taken to be $\mathbf{x} = (0,0,1/\sqrt{12},1/\sqrt{12},1/\sqrt{12})$. In \Fig{fig:timeevol_ada}, we similarly show the time-evolution of the normalized photon number $|\alpha(t)|^2$ for the same initial state and parameters. These figures clearly show the exceedingly slow approach towards the stable steady state, in agreement with the predictions from Ref.\cite{Bhaseen12}. 

In order to obtain  the characteristic time-scales governing the approach to a fixed point, $\bar\mathbf{x}(t) \equiv \bar\mathbf{x}$, when feedback is applied, we consider a small perturbation, $\mathbf{y}(t) = \mathbf{x}(t) - \bar\mathbf{x}$, and linearize the equations of motion around the solution. By using an Ansatz $\mathbf{y}(t) = \exp(\lambda t)\mathbf{y}_0$, we can then derive a characteristic equation for $\lambda$. Each solution $\lambda$ corresponds to a characteristic (inverse) time-scale for the dynamics close to the steady state. In the presence of feedback, $k,\tau > 0$, there are in fact an infinite number of solutions $\lambda_k$. However, a crucial result in the analysis of delay differential equations is that there are only a finite number in any real half-plane $\Re\lambda > \sigma, \, \sigma \in \mathbb{R}$ \cite{Bellen13}. Thus it becomes feasible to find the slowest $\lambda_k$ that ultimately governs the time-scale for approaching or leaving a fixed point. The details of a stability analysis for \eq{eq:semiclass} are given in \ref{sect:app_stability}.

A steady state solution $\bar\mathbf{x}$ is stable only if all $\lambda_k$ have negative real parts, i.e., $\Re \lambda_k  < 0$. The time-scale for the approach to a stable steady state solution is thus governed by the eigenvalue with real part closest to zero, which we will denote by $\lambda_1$. The key to our control scheme is that the eigenvalues can be manipulated through the variation of $k$ and $\tau$. In particular, both the magnitude and sign of $\Re \lambda_1 $ can be changed. This opens up the possibility of changing the emergent time-scales, as well as qualitatively changing the phase diagram of the system, by changing the stability of a steady state.

\begin{figure}
\vspace{100mm}
\begin{picture}(0,0)(-20,0)   
\includegraphics{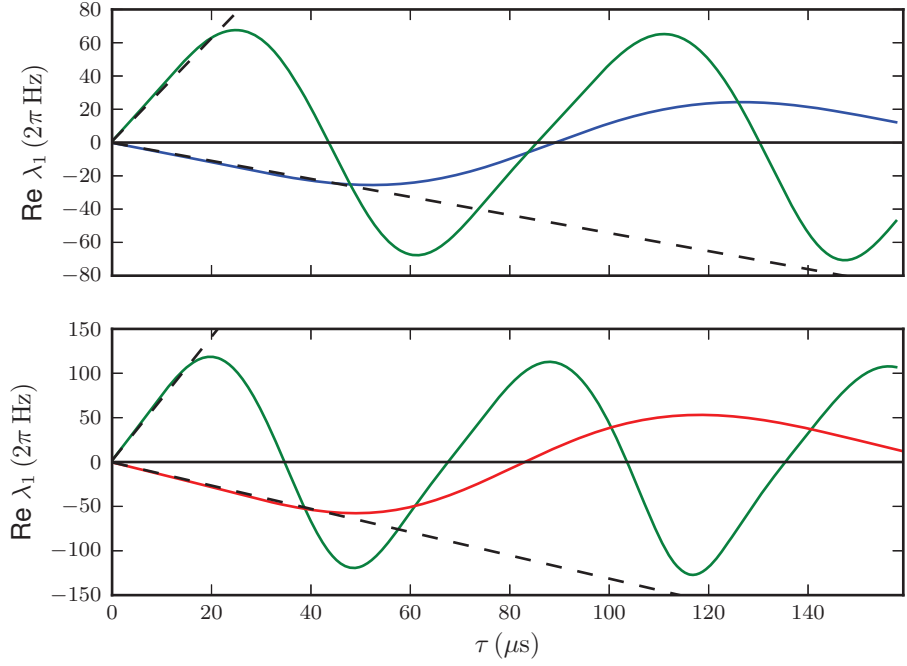}
\end{picture}
  \caption{\label{fig:stability} The root of \eq{eq:char} with the real part closest to zero varies as the delay $\tau$ is changed. This sets the characteristic relaxation time $1/\Re \lambda_1$. {\bf Top panel}: $g/g_c = 0.74$. The blue line is from a linearization around the normal phase, $\mathbf{x}^\Downarrow$ \eq{eq:norm}, and the green line is for the inverted phase, $\mathbf{x}^\Uparrow$ \eq{eq:inv}. {\bf Bottom panel}: $g/g_c = 1.1$. The green line is for the inverted phase, and the red line is for the super-radiant phase, $\mathbf{x}^\text{SR}$ \eq{eq:SR}. The normal phase is not shown for this value of $g$, but has a value of $\Re \lambda_1 \simeq 3.5 \cdot 2\pi$ kHz throughout (and is thus unstable). The dashed lines show the corresponding linear approximations given in (\ref{eq:lambda_approx_a},\ref{eq:lambda_approx_b}). For both panels, $k$ is set to $\kappa/2$, and all other parameters are given in \eq{eq:params}.}
\end{figure}

We can find $\lambda_1$ numerically by solving a transcendental characteristic equation, given in \eq{eq:char}. In \Fig{fig:stability}, we plot $\Re \lambda_1$ as a function of $\tau$ for $k=\kappa/2$, $g/g_c = 0.74$ and $1.1$, while the other parameters are kept as in  \eq{eq:params}. For $g/g_c = 0.74$, we linearize around the normal phase, $\mathbf{x}^\Downarrow$ \eq{eq:norm}, and the inverted phase $\mathbf{x}^\Uparrow$ \eq{eq:inv}, which are always valid fixed points. For $g/g_c = 1.1$ we also linearize around the super-radiant fixed point, $\mathbf{x}^\text{SR}$ \eq{eq:SR}. This analysis can be used to find optimal values for $\tau$ close to the steady state. We observe that a minimum value for $\Re \lambda_1$ is reached for a delay of around $\tau \simeq 50\,\mu$s, for both the normal phase when $g/g_c=0.74$ and the super-radiant phase when $g/g_c=1.1$. The value of $\Re\lambda_1$ is, without feedback ($\tau = 0$), roughly  $-0.14 \cdot 2\pi$ Hz and $-0.35 \cdot 2\pi$ Hz, respectively, for the two cases. In comparison, at the first minima, with $\tau=52\,\mu$s and $\tau=50\,\mu$s, the values are $-26 \cdot 2\pi$ and $-58\cdot 2\pi$ Hz, i.e., two orders of magnitude larger.

In \Fig{fig:timeevol} and \Fig{fig:timeevol_ada} the time-evolution of the collective inversion, $j_z(t)$, and the normalized photon number, $|\alpha(t)|^2$, from an initial state $\mathbf{x}=(0,0,1/\sqrt{12},1/\sqrt{12},1/\sqrt{12})^T$, is shown for various values of $\tau$, with $k$ set to $\kappa/2$, and either $g/g_c=0.74$ or $g/g_c=1.1$. The other parameters are again as given by \eq{eq:params}. With $\tau=50\,\mu$s the steady state is reached after $\sim 20$ ms, a dramatic improvement when compared to the dynamics without feedback, for which the relaxation takes several seconds. These figures also show the possibility of qualitatively changing the phase diagram by choosing a value of $\tau$ that de-stabilizes a fixed point. Note that these results were found by numerically integrating \eq{eq:semiclass}, using an integrator designed for delay differential equations \cite{Flunkert09}.

We note that significant improvement can also be achieved for smaller delay-times than those used in \Fig{fig:timeevol} and \Fig{fig:timeevol_ada}. We find an approximate expression for $\lambda_1$, valid for small $\lambda_1\tau$, in the form $\lambda_1 \simeq \lambda^{(0)} + \lambda^{(1)}$ with
\numparts
  \begin{eqnarray}
    \lambda^{(0)} =& i\sqrt{\tilde{\omega}_0^2 +\frac{4g\tilde{\omega}_0 \bar{x}_1\bar{j}_x}{\bar{j}_z} + \frac{2\tilde{\omega}\tilde{\omega}_0\left|2g\bar{j}_z-U\bar{\alpha}\bar{j}_x\right|^2}{\left(\kappa^2+\tilde{\omega}^2\right)\bar{j}_z}}\label{eq:lambda_approx_a}\, , 
\end{eqnarray}
and
\begin{eqnarray}
    \lambda^{(1)} =& \kappa(1+k\tau) \frac{2\tilde{\omega}\tilde{\omega}_0\left|2g\bar{j}_z-U\bar{\alpha}\bar{j}_x\right|^2}{\left(\kappa^2+\tilde{\omega}^2\right)^2\bar{j}_z}\label{eq:lambda_approx_b}\, .
\end{eqnarray}
\endnumparts
Here we have introduced $\tilde{\omega} \equiv \omega + U \bar{j}_z$ and $\tilde{\omega}_{0} \equiv \omega_0 + U(\bar{x}_1^{2}+\bar{x}_2^{2}) \equiv \omega_0 + U|\bar{\alpha}|^2$, where $\bar{\alpha} \equiv \bar{x}_1 + i\bar{x}_2$. More details are presented in \ref{sect:app_stability}. For the parameters we use, we find that $\lambda^{(0)}$ is imaginary. The characteristic time-scale is for small $\lambda_1 \tau$ thus set by $\Re\lambda_1 \simeq \lambda^{(1)}$. We plot these approximate solutions together with the exact numerical solutions in \Fig{fig:stability}, shown as dashed lines. We see that this linear approximation of $\Re\lambda_1$  captures the small $\tau$ behavior very well.

\begin{figure}
%
%
\vspace{130mm}
\begin{picture}(0,0)(-20,0)   
\includegraphics{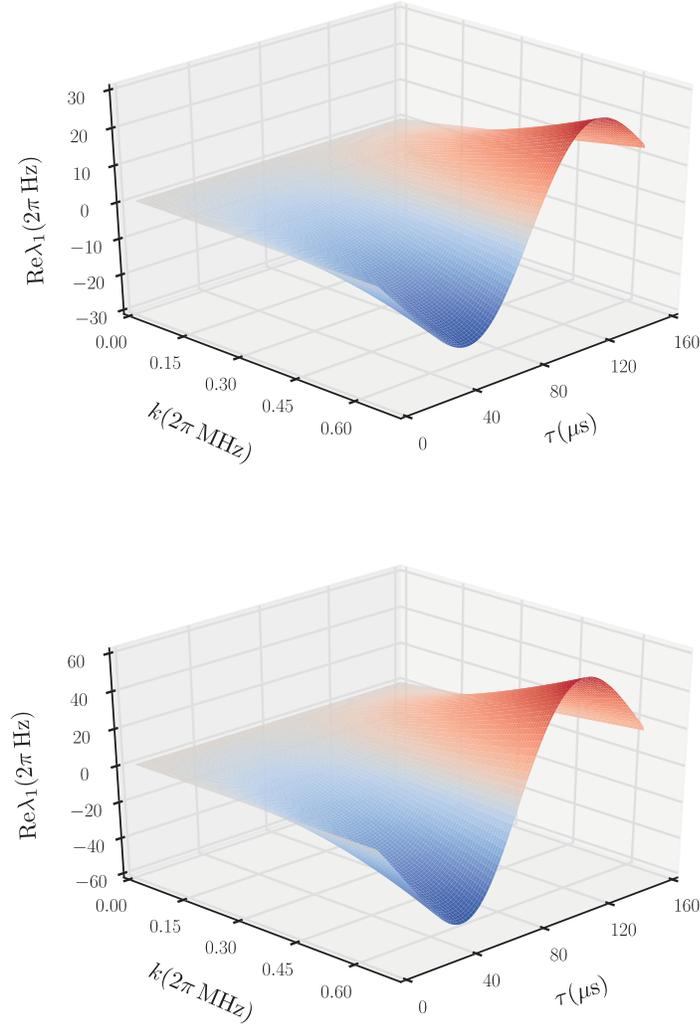}
\end{picture}
  \caption{\label{fig:stability_surf} $\Re \lambda_1$ as a function of $(k,\tau)$. {\bf Top}: Results from a linearization around the normal phase, $\mathbf{x}^\Downarrow$ \eq{eq:norm}, with $g/g_c=0.74$. {\bf Bottom}: Similarly for the super-radiant phase, $\mathbf{x}^\text{SR}$ \eq{eq:SR}, and $g/g_c=1.1$. The other parameters are given in \eq{eq:params}.}
\end{figure}

So far, we have not accounted for any loss in the feedback loop, and considered only the ideal case $k=\kappa/2$ in our numerical results. However, good results are also achieved for smaller $k$, as already indicated by the approximate linear dependence in \eq{eq:lambda_approx_b}. In \Fig{fig:stability_surf}, we show $\Re \lambda_1$ as a function of $k$ and $\tau$, for linearization around two steady states: the normal phase, for $g/g_c=0.74$, and the super-radiant phase, for $g/g_c=1.1$. The other parameters are as before \eq{eq:params}. This shows that great improvements in the convergence are also possible with significant loss in the feedback loop. For the choice $k= 0.1\cdot \kappa/2$ (90\% loss) we, e.g.,  still expect an order of magnitude improvement in the relaxation time for the parameters considered.

\section{\label{sect:sweep}Finite-time experiments and unexplored regions of the phase diagram}

We have illustrated the possibility of very slow time-scales associated with the generalized Dicke model when realistic parameters are used, by considering the time-evolution of an initial state, chosen far from equilibrium (\Fig{fig:timeevol} and \Fig{fig:timeevol_ada}).  It is, however, important to note that in a typical experiment \cite{Baumann10,Baumann11,Brennecke13} the system is \emph{prepared} in the normal phase for some $g<g_c$, before gradually ramping up $g$ to a value beyond $g_c$. Due to the continuous nature of the phase transition at $g_c$, the change in steady state is gradual enough that the system can react to the small rate of change. As the experiments have shown, as well as the theoretical modelling in Ref.\cite{Bhaseen12}, the measured photon intensity agrees well with the semi-classical steady state value.

We will here follow the approach in Ref.\cite{Bhaseen12} to emulate a finite-time experiment. We will start with the system prepared close to the normal phase, and then ramp up $g(t) \sim \sqrt{t}$ to a value beyond the critical point. Specifically, we will choose $g(t) = \sqrt{t/t_0} \cdot 1.5 g_c$. To emulate the effect of quantum fluctuations, we  prepare initial state of the system close to the normal phase, with $j_x = j_y = 1/\sqrt{N}$, and $N=10^5$.

\begin{figure}
%
\vspace{100mm}
\begin{picture}(0,0)(-20,0)   
\includegraphics{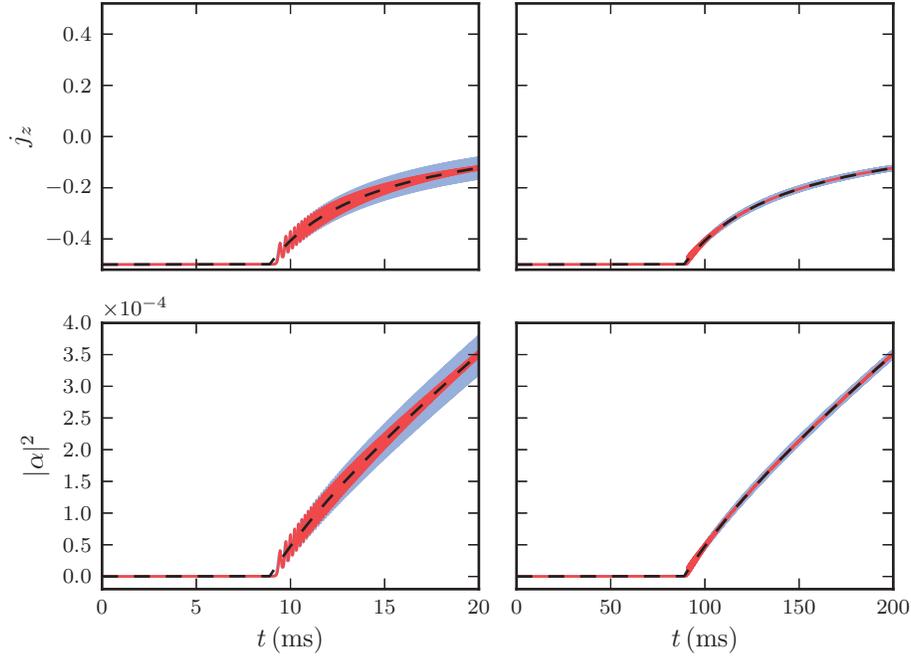}
\end{picture}
  \caption{\label{fig:ramp1} $j_z(t)$ and $|\alpha(t)|^2$ found by ramping up $g$ according to $g(t) = \sqrt{t/t_0} \cdot 1.5 g_c$. The {\bf left panels} are for $t_0=20$~ms, and the {\bf right panels} are for $t_0=200$~ms. The blue shaded regions show the (rapidly oscillating) time-evolution without feedback. The red shaded regions are similarly for $(k,\tau) = (\kappa/2,16\,\mu\text{s})$. The dashed lines show the semi-classical fixed point values. The other parameters are as in \eq{eq:params}.}
\end{figure}

We now envision an experiment where the goal is to reach the steady state well beyond the critical point, at $g/g_c = 1.5$. For this purpose, we perform a linearization around this steady state, and find a good value for $\tau$, as was illiustrated in \Fig{fig:stability}. We use, again, $k=\kappa/2$, and the other parameters as in \eq{eq:params}. We then find a minimum for $\Re \lambda_1$ around $\tau = 16\,\mu$s. In \Fig{fig:ramp1}, we show the time-evolution of the collective inversion and the photon number as $g(t)$ is ramped up, and compare the situation with and without feedback. Furthermore, we compare two sweeps, one with $t_0=20$ ms, and one with $t_0=200$ ms. We observe that the system, both with and without feedback, responds to the change across the critical point, and closely follows the adiabatic evolution according to the exact steady state values. We, however,  note that in the case with feedback, the fluctuations are much smaller. A fully quantum treatment of fluctuations will be given in the next section.

\begin{figure}
%
\vspace{100mm}
\begin{picture}(0,0)(-20,0)   
\includegraphics{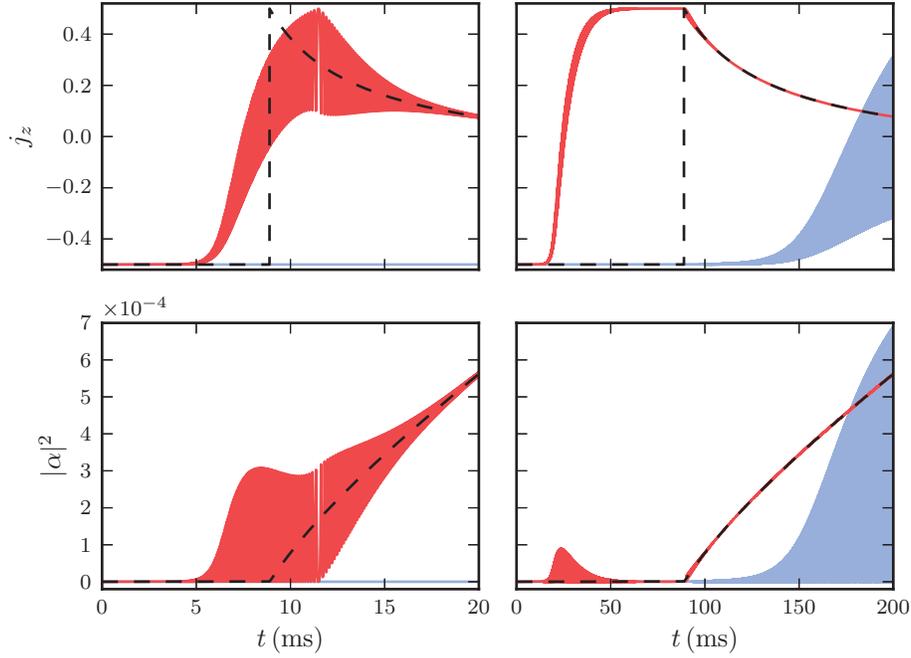}
\end{picture}
  \caption{\label{fig:ramp2} $j_z(t)$ and $|\alpha(t)|^2$ found by ramping up $g$ according to $g(t) = \sqrt{t/t_0} \cdot 1.5 g_c$. The {\bf left panels} are for $t_0=20$~ms, and the {\bf right panels} are for $t_0=200$~ms.  The blue shaded regions show the (highly oscillating) time-evolution without feedback. The red shaded regions are similarly for $(k,\tau) = (\kappa/2,16\,\mu\text{s})$. The dashed lines show the semi-classical fixed point values. The other parameters are as in \eq{eq:params}, except for $\omega$ which is set to $\omega = -10.0 \cdot 2\pi$ MHz.}
\end{figure}

One of the main findings in Ref.\cite{Bhaseen12} was, however, that there are regions of the phase diagram where the situation is far more problematic. This was particularly found to be the case for negative $\omega$ ($\omega < U/2 <0$), where emulation of finite time experiments with sweeps up to 200 ms were not able to approximate the super-radiant steady state. This region of parameter space has also not been explored experimentally. 
Here the normal phase is unstable below threshold, but the time-scale for leaving the normal phase is extremely slow. If the system is prepared close to the normal phase, it is therefore essentially meta-stable on typical experimental run-times. However, it will be far from steady state when the experiment hits threshold, $g=g_c$, and the system is unable to ``respond'' to the super-radiant phase transition on an adequate time-scale. 
In \Fig{fig:ramp2}, we exhibit finite-time sweeps with $t_0=20$ ms and 200 ms, but now for $\omega = -10.0\cdot 2\pi$ MHz. The other parameters are kept as in \eq{eq:params}, and we compare the case with no feedback ($k=0$) to the case with $(k,\tau)=(\kappa/2,16\,\mu\text{s})$ as before. We see that, without feedback, the phase transition is not visible due to the slow response of the system. In other words, the time-evolution is far from being adiabatic. The results with feedback, on the other hand, are in this case quite interesting. At this value of $\tau$, the system responds rapidly, and does a relatively fast switch from the normal to the inverted phase, before hitting threshold. The evolution subsequently adapts well to the super-radiant phase transition across $g_c$, and for both sweep times the target steady state at $g/g_c=1.5$ is reached to a very good approximation.


\section{\label{sect:quantumfluct}Quantum Fluctuations}

So far, we have considered the semi-classical amplitudes of the relevant observables. It is of importance to consider how quantum fluctuations are influenced by our feedback control scheme. We consider the thermodynamic $N\to\infty$ limit, and follow the treatment in Refs. \cite{Emary03a,Emary03b,Dimer07} by introducing a Holstein-Primakoff representation for the collective spin, i.e.
\numparts
\begin{eqnarray}
  \hat{J}_z = \hat{b}\dagg \hat{b} - \frac{N}{2}\, ,
\end{eqnarray}
and
\begin{eqnarray}
  \hat{J}_+ =& \hat{b}\dagg\sqrt{N-\hat{b}\dagg \hat{b}} = \left(\hat{J}_-\right) \dagg \, ,
\end{eqnarray}
\endnumparts
where $\hat{J}_\pm = \hat{J}_x \pm i \hat{J}_y$, and $\hat{b}$ ($\hat{b}\dagg$) are bosonic annihilation (creation) operators satisfying $[\hat{b},\hat{b}\dagg]=1$. We next introduce fluctuation operators by expanding the fields around their semi-classical amplitudes, $\hat{a} \equiv \langle\hat{a}\rangle + \delta\hat{a}, \hat{b} \equiv \langle\hat{b}\rangle + \delta\hat{b}$, where we use $\langle\hat{b}\rangle = \pm \sqrt{N/2+\langle\hat{J}_z\rangle}$. For the normal phase we have $\langle\hat{a}\rangle = \langle\hat{b}\rangle = 0$, whereas in the super-radiant phase we have $\langle\hat{a}\rangle/\sqrt{N} = \alpha^\text{SR}, \langle\hat{b}\rangle/\sqrt{N} = \pm\sqrt{1/2+j_z^\text{SR}}$, where $\alpha^\text{SR},j_z^\text{SR}$ are given in \eq{eq:SR}. We observe  that a positive choice for $\langle\hat{b}\rangle$ corresponds to a negative choice for $\langle\hat{a}\rangle$, and vice versa. In the following, we will just consider the positive choice for $\langle\hat{b}\rangle$, as the calculation with the other choice is essentially identical, with a few changes in sign.


After a lengthy calculation, one finds that the fluctuations satisfy the following equations of motion, in the limit $N\to\infty$ (see \ref{sect:app_fluct}):
\numparts
\begin{eqnarray}
  \delta\dot{\hat{a}} = i[\hat{H}',\delta \hat{a}] - \kappa \delta\hat{a} 
                       +k\left(\delta\hat{a}(t-\tau)-\delta\hat{a}(t)\right) -\sqrt{2\kappa}\hat{a}_\text{in}(t) \label{eq:fluct_time_a}\, , 
\end{eqnarray}
and
\begin{eqnarray}
  \delta\dot{\hat{b}} = i[\hat{H}',\delta \hat{b}] \label{eq:fluct_time_b}\,  ,
\end{eqnarray}
\endnumparts
with
\begin{eqnarray}\label{eq:H_HP}
  \hat{H}' =& \omega_a\delta\hat{a}\dagg\delta\hat{a} + \omega_b\delta\hat{b}\dagg\delta\hat{b} + \lambda_1(\delta\hat{a}+\delta\hat{a}\dagg)(\delta\hat{b}+\delta\hat{b}\dagg) \nonumber\\
      &+ i\lambda_2(\delta\hat{a}-\delta\hat{a}\dagg)(\delta\hat{b}+\delta\hat{b}\dagg) + \frac{\chi}{4}(\delta\hat{b}+\delta\hat{b}\dagg)^2\, ,
\end{eqnarray}
and where
\numparts
\begin{eqnarray}
  \omega_a =& \omega + U\bar{j}_z\, ,\\
  \omega_b =& \omega_0+\frac{4g^2\omega_a}{\omega_a^2+\kappa^2}\left(\half+\bar{j}_z\right)+U|\bar{\alpha}|^2\, ,\\
  \lambda_1 =& -\frac{2g\bar{j}_z}{\sqrt{\half-\bar{j}_z}}-\frac{2gU\omega_a}{\omega_a^2+\kappa^2}\left(\half+\bar{j}_z\right)\sqrt{\half-\bar{j}_z}\, ,\\
  \lambda_2 =& \frac{2gU\kappa}{\omega_a^2+\kappa^2}\left(\half+\bar{j}_z\right)\sqrt{\half-\bar{j}_z}\, ,\\
  \chi =& \frac{4g^2\omega_a}{\omega_a^2+\kappa^2} \frac{\left(\half+\bar{j}_z\right)\left(\frac{3}{2}-\bar{j}_z\right)}{\half-\bar{j}_z}\, .
\end{eqnarray}
\endnumparts
These expressions are valid in both the normal and the super-radiant phase, where $\bar{\alpha}$ and $\bar{j}_z$ refer to the semi-classical amplitudes in the respective phases. 

The set of euqations (\ref{eq:fluct_time_a}--\ref{eq:fluct_time_b}) are linear, and as was done in Ref.\cite{Dimer07}, we solve these equations by making use of Fourier-transform techniques as explained in more detail in \ref{sect:app_fourier}. The mean fluctuations in the intra-cavity photon number in steady state can then be computed through
\begin{eqnarray}
  \braket{\delta \hat{a}\dagg \delta \hat{a}}_\text{ss} = \frac{1}{2\pi} \int_{-\infty}^\infty \int_{-\infty}^\infty \braket{\delta \tilde{a}\dagg(\nu)\delta \tilde{a}(\nu')} \dx{\nu}\dx{\nu'}\, ,
\end{eqnarray}
where $\delta\tilde{a}(\nu)$ is the Fourier transformed cavity field fluctuation.

\begin{figure}[h]
%
\vspace{100mm}
\begin{picture}(0,0)(-20,0)   
\includegraphics{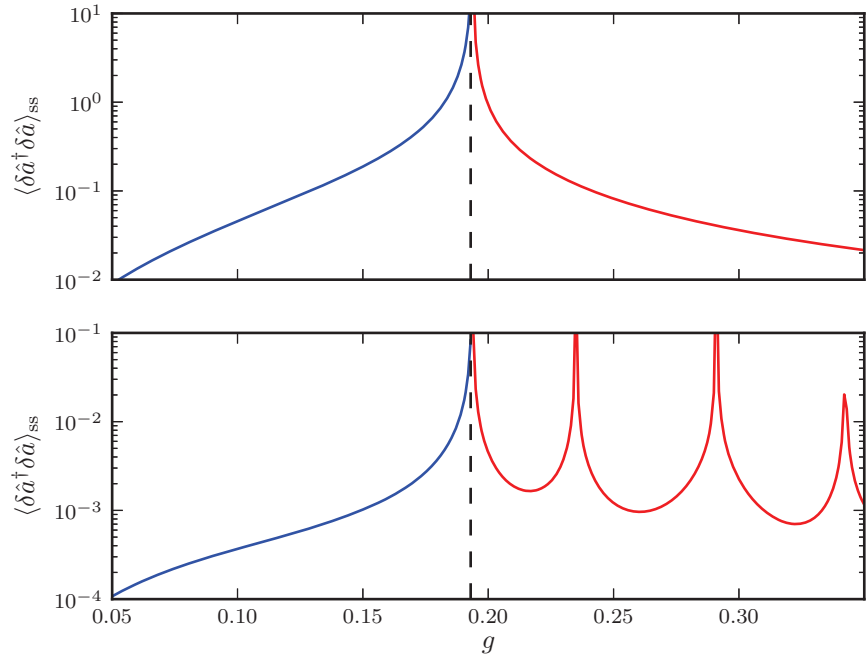}
\end{picture}
  \caption{\label{fig:fluct} Divergence of quantum fluctuations upon approaching the critical point. The blue lines show the normal phase, and the red lines show the super-radiant phase. The dashed vertical lines show the critical value for $g$. {\bf Top panel}: Without feedback ($k=0$). {\bf Bottom panel}: $(k,\tau) = (\kappa/2,50\,\mu\text{s})$. For both panels, all other parameters are given in \eq{eq:params}.}
\end{figure}

We solve this integral numerically, while varying $g$, to investigate the diverging fluctuations upon approaching the critical value $g_c$. In \Fig{fig:fluct}, we compare the case with no feedback ($k=0$) to the case with $(k,\tau) = (\kappa/2,50\,\mu\text{s})$. We see that the feedback can significantly influence the size of the quantum fluctuations. The change of about two orders of magnitude is consistent with the results from the previous sections. Interestingly, we observe that a series of new instabilities develop in the super-radiant phase for increasing $g$ when feedback is applied. We also note that the scaling of $\left(g_c/g\right)\sqrt{\braket{\delta \hat{a}\dagg \delta \hat{a}}_\text{ss}}$ with $|1-g/g_c|$, as $g\to g_c$, is found to be the same in the two panels of \Fig{fig:fluct}, corresponding to a universal ``photon flux exponent'' of 1.0 \cite{Nagy11,Brennecke13}. Fluctuations in the $\hat{b}$-mode, $\langle \delta \hat{b}\dagg \delta\hat{b} \rangle_\text{ss}$, show similar divergences at the same values of $g$.

\section{\label{sect:conc}Conclusions and outlook}

We have modelled Pyragas' time-delay auto-synchronization feedback control applied to a quantum system. For a topical cavity QED many-body system, the Dicke model as realized in recent experiments, we have investigated  in what manner a  feedback force can influence the characteristic time-scales governing the relaxation of the system as well as the characteristic size of quantum fluctuations. With optimal feedback times, the relaxation-time can be reduced by two orders of magnitude, with a corresponding decrease in quantum fluctuations. Even with significant loss in the feedback loop (90\% loss), one expects an order of magnitude improvement. The  scheme put forward also offers the possibility of changing the stability of long time attractors, thus qualitatively changing the phase diagram of the system. A fixed point might, e.g.,  be de-stabilized at a critical value of the delay-time $\tau$ and  thus inducing a novel feedback-driven phase transition. 

Although we have focused specifically on the Dicke model in this paper, and particularly on how to reduce the relaxation time, we believe the scheme could be applied to a variety of topical CQED systems. It might, e.g.,  be interesting to consider systems with optical bistability, where the scheme may be used to change the stability of the fixed points. In other systems, it may also be of interest to consider how the feedback control can \emph{slow down} the dynamics, instead of making the convergence more rapid, if the goal is to preserve quantum information in an initial state. In general, the scheme presented in the present paper offers a novel and \emph{non-invasive} way to control the overall time-scale governing the dynamics of open quantum systems, where non-invasive here means that the positions of fixed points in parameters space remain unchanged, only their relative stability is changed.

\ack
The authors are grateful for the hospitality shown at the University of Auckland (ALG)  and at KITP, the University of California at Santa Barbara (B.-S.S), when the present paper was in progress. ALG thanks Ferdinand Brennecke and Anup Purewal for helpful discussions. The authors acknowledge the contribution of NeSI high-performance computing facilities to the results of this research. New Zealand's national facilities are provided by the NZ eScience Infrastructure and funded jointly by NeSI's collaborator institutions and through the Ministry of Science \& Innovation's Research Infrastructure program. URL http://www.nesi.org.nz.
This work has been supported in part by the Norwegian University of Science and Technology (NTNU) and,
for one of the authors (B.-S.S), also by the Norwegian Research Council under contract NFR 191564/V30, ''Complex
Systems and Soft Materials'', and in part by the National Science Foundation under Grant. No. NSF
PHY11-25915.

\appendix

\section{\label{sect:app_stability}Stability and characteristic time-scales}

Here we summarize the semi-classical stability analysis, leading to the characteristic time-scale governing the time-evolution close to a fixed point. 
For notational convenience, we will write $\mathbf{x} \equiv (x_1,x_2,j_x,j_y,j_z)$ and define a real vector-valued function $\mathbf{f}$ such that \eq{eq:semiclass} can be written in the form
\begin{eqnarray}\label{eq:diffeq}
  \frac{d\mathbf{x}(t)}{dt}= \mathbf{f}(\mathbf{x}) + k\mathbf{B}\cdot(\mathbf{x}(t-\tau)-\mathbf{x}(t))\, ,
\end{eqnarray}
where the matrix $\mathbf{B} \equiv \text{diag}(1,1,0,0,0)$.

We now consider a small perturbation, $\mathbf{y}(t) \equiv \mathbf{x}(t) - \bar{\mathbf{x}}(t)$, from a solution $\bar{\mathbf{x}}(t)$ to \eq{eq:diffeq}, and linearize the equations of motion, \emph{i.e.},
\begin{eqnarray}\label{eq:linear}
  \frac{d\mathbf{y}(t)}{dt}= \mathbf{A}(t)\cdot \mathbf{y}(t) + k\mathbf{B}\cdot\left(\mathbf{y}(t-\tau) - \mathbf{y}(t)\right)\, ,
\end{eqnarray}
where $A_{ij} = \partial{f_i}/\partial{x_j}$. If $\bar\mathbf{x}(t) \equiv \bar\mathbf{x} = (\bar{x}_1,\bar{x}_2,\bar{j}_x,\bar{j}_y,\bar{j}_z)$ is a steady state solution, the matrix $\mathbf{A}(t) \equiv \mathbf{A}$ is time-independent. 

Before proceeding, it is convenient to eliminate one of the spin variables from \eq{eq:linear} by using the constraint \eq{eq:constraint}. By differentiating this constraint, and using the Ansatz $\mathbf{y}(t) = \exp(\lambda t)\mathbf{y}_0$, we eliminate $j_z$, leading to a linear set of equations for $\mathbf{z}(t) \equiv (x_1(t)-\bar{x}_1,x_2(t)-\bar{x}_2,j_x(t)-\bar{j}_x,j_y(t)-\bar{j}_y)$, i.e., 
\begin{eqnarray}\label{eq:linear2}
  \frac{d\mathbf{z}(t)}{dt} = \mathbf{A}'\cdot \mathbf{z}(t) + k\mathbf{B}'\cdot\left(\mathbf{z}(t-\tau) - \mathbf{z}(t)\right),
\end{eqnarray}
where $\mathbf{B}'\equiv\text{diag}(1,1,0,0)$ and the matrix $\mathbf{A}'$ reads
\begin{eqnarray}
  \fl
  \mathbf{A}' = \left(\begin{array}{cccc}
      -\kappa & \tilde{\omega} & -U\bar{x}_2\bar{j}_x/\bar{j}_z & -U\bar{x}_2\bar{j}_y/\bar{j}_z \\
      -\tilde{\omega} & -\kappa & -2g + U\bar{x}_1\bar{j}_x/\bar{j}_z & U\bar{x}_1\bar{j}_y/\bar{j}_z \\
      -2U\bar{x}_1 \bar{j}_y & -2U\bar{x}_2 \bar{j}_y & 0 & -\tilde{\omega}_0 \\
      2U\bar{x}_1 \bar{j}_x - 4g\bar{j}_z & 2U\bar{x}_2 \bar{j}_x & \tilde{\omega}_0 + 4g\bar{x}_1\bar{j}_x/\bar{j}_z & 4g\bar{x}_1\bar{j}_y/\bar{j}_z 
  \end{array}\right)\, .
\end{eqnarray}
Here we have introduced the notation $\tilde{\omega} \equiv \omega + U \bar{j}_z$ and $\tilde{\omega}_{0} \equiv \omega_0 + U(\bar{x}_1^{2}+\bar{x}_2^{2}) \equiv \omega_0 + U|\bar{\alpha}|^2$, where $\bar{\alpha} \equiv \bar{x}_1 + i\bar{x}_2$. Each solution to the characteristic equation for $\lambda$,
\begin{eqnarray}\label{eq:char}
  \det \Delta(\lambda) = 0\, ,
\end{eqnarray}
where
\begin{eqnarray}
  \Delta(\lambda) = \lambda \mathbf{I}_4 - \mathbf{A}' - k\left(\e^{-\lambda \tau}-1\right)\mathbf{B}'\, ,
\end{eqnarray}
corresponds to a characteristic (inverse) time-scale for the dynamics close to the steady state. Here $\mathbf{I}_4$ is the 
$4\times 4$ identity matrix. For all the fixed points we are interested in, we have $j_y=0$, which simplifies \eq{eq:char}. Explicitly, we find for $j_y=0$ that
\begin{eqnarray}\label{eq:char2}
  \det \Delta(\lambda)  = \left[\tilde{\omega}^2 + \left(\lambda + \kappa + k(1-\e^{-\lambda\tau})\right)^2\right] \nonumber\\
                         \times \left(\lambda^2+\tilde{\omega}_{0}^2 + \frac{4g\tilde{\omega}_0 \bar{x}_1\bar{j}_x}{\bar{j}_z}\right)
                         +\frac{2\tilde{\omega}\tilde{\omega}_0}{\bar{j}_z}\left|2g\bar{j}_z-U\bar{\alpha}\bar{j}_x\right|^2 = 0\, . 
\end{eqnarray}
\noindent The transcendental equation (\ref{eq:char2}) has an infinite number of roots when $k,\tau > 0$. However, a crucial result in the analysis of delay differential equations is that there are only a finite number of roots in any real half plane, $\Re\lambda > \sigma, \sigma \in \mathbb{R}$ (see e.g. Ref.\cite{Bellen13}). Thus it becomes possible to find the smallest root, which we denote by $\lambda_1$, that ultimately governs the time-scale for approaching or leaving a fixed point. In \Fig{fig:stability} and \Fig{fig:stability_surf}, $\lambda_1$ was found numerically for varying $\tau$ and $k$ and different fixed points $\bar{\bf x}$. This was done using a MATLAB tool for analyzing linear delay differential equations \cite{ddebiftool}.

To get a handle on the behavior of $\lambda_1$ for small delays, we can approximate the characteristic equation \eq{eq:char2}, for small $\lambda\tau$, using $1-\exp(-\lambda\tau) \simeq \lambda\tau$. We further anticipate $\lambda \simeq \lambda^{(0)} + \lambda^{(1)}$ where $\lambda^{(0)} \sim \tilde{\omega}_0 \ll \kappa$, $\lambda^{(1)} \sim \tilde{\omega}_0^2 \ll \tilde{\omega}_0$ and $\bar{x}_1 \bar{j}_x/\bar{j}_z \sim \tilde{\omega}_0$. Using this, we find
%
%
  \begin{eqnarray}
    \lambda^{(0)} =i\sqrt{\tilde{\omega}_0^2 +\frac{4g\tilde{\omega}_0 \bar{x}_1\bar{j}_x}{\bar{j}_z} + \frac{2\tilde{\omega}\tilde{\omega}_0\left|2g\bar{j}_z-U\bar{\alpha}\bar{j}_x\right|^2}{\left(\kappa^2+\tilde{\omega}^2\right)\bar{j}_z}}\, ,
\end{eqnarray}
and
\begin{eqnarray}
    \lambda^{(1)} = \kappa(1+k\tau) \frac{2\tilde{\omega}\tilde{\omega}_0\left|2g\bar{j}_z-U\bar{\alpha}\bar{j}_x\right|^2}{\left(\kappa^2+\tilde{\omega}^2\right)^2\bar{j}_z}\, .
\end{eqnarray}
%
%
For the parameters we use, we find that $\lambda^{(0)}$ is imaginary. The (inverse) characteristic time-scale is, for small $\lambda \tau$, thus set by $\Re\lambda_1 \simeq \lambda^{(1)}$.

\section{\label{sect:app_fluct}Quantum fluctuations in the thermodynamic limit}

Here we outline in some more detail the calculations leading to Eqs. (\ref{eq:fluct_time_a}), (\ref{eq:fluct_time_b}) and (\ref{eq:H_HP}). After introducing the Holstein-Primakoff representation, i.e., 
%
%
\begin{eqnarray}
  \hat{J}_z =& \hat{b}\dagg \hat{b} - \frac{N}{2}\, ,\\
  \hat{J}_+ =& \hat{b}\dagg\sqrt{N-\hat{b}\dagg \hat{b}} = \left(\hat{J}_-\right)\dagg \, ,
\end{eqnarray}
%
%
and neglecting constant energy terms, we have that the Dicke Hamiltonian \eq{eq:H_Dicke} can be rewritten in the form
\begin{eqnarray}
  \hat{H} =& \omega_0 \hat{b}\dagg \hat{b} + \left(\omega-\frac{U}{2}\right)\hat{a}\dagg \hat{a} \nonumber \\
           & +\frac{g}{\sqrt{N}}\left(\sqrt{N-\hat{b}\dagg \hat{b}}~\hat{b} + \hat{b}\dagg\sqrt{N-\hat{b}\dagg \hat{b}}\right)\left(\hat{a}+\hat{a}\dagg\right) \nonumber \\
           &+ \frac{U}{N}\hat{b}\dagg \hat{b} \hat{a}\dagg \hat{a}\, .
\end{eqnarray}
Next, we expand the fields as $\hat{a} = a_0 + \delta\hat{a}, \hat{b} = b_0 + \delta\hat{b}$, where $a_0 \in \mathbb{C}, b_0 \in \mathbb{R}$, and write:
\begin{eqnarray}\label{eq:H_A1}
  \hat{H} =& \omega_0 (b_0+\delta \hat{b}\dagg)(b_0+\delta\hat{b}) + \left(\omega-\frac{U}{2}\right)(a_0\conj+\delta\hat{a}\dagg)(a_0+\delta\hat{a}) \nonumber\\
           & +\frac{g}{\sqrt{N}}\left(\hat\xi(b_0+\delta\hat{b}) + (b_0+\delta\hat{b}\dagg)\hat\xi\right)
            \left(a_0+\delta\hat{a}+a_0\conj+\delta\hat{a}\dagg\right) \nonumber \\
           &+ \frac{U}{N}(b_0+\delta\hat{b}\dagg)(b_0+ \delta\hat{b})(a_0\conj+\delta\hat{a}\dagg)(a_0+\delta\hat{a})\, ,
\end{eqnarray}
where
\begin{eqnarray}
  \hat{\xi} = \sqrt{N-(b_0+\delta\hat{b}\dagg)(b_0+\delta\hat{b})}\, .
\end{eqnarray}
Let us first consider the normal phase, where we set $a_0=b_0=0$, before considering the more involved super-radiant phase.

\subsection{The normal phase}

In this case, we have that \eq{eq:H_A1} becomes
\begin{eqnarray}
  \hat{H} =& \omega_0 \delta \hat{b}\dagg\delta\hat{b} + (\omega-U/2)\delta\hat{a}\dagg\delta\hat{a} \nonumber \\
            +\frac{g}{\sqrt{N}}&\left(\hat\xi\delta\hat{b} + \delta\hat{b}\dagg\hat\xi\right)\left(\delta\hat{a}+\delta\hat{a}\dagg\right) 
           + \frac{U}{N}\delta\hat{b}\dagg\delta\hat{b}\delta\hat{a}\dagg\delta\hat{a}\, .
\end{eqnarray}
Next, we make use of the approximation
\begin{eqnarray}
  \hat{\xi} = \sqrt{N-\delta\hat{b}\dagg\delta\hat{b}} \simeq \sqrt{N}\, ,
\end{eqnarray}
and by inserting this into the Hamiltonian above, and neglecting the term proportional to $U/N$ that vanish in the limit $N\to\infty$, we obtain an effective Hamiltonian for the normal phase:
\begin{eqnarray}
  \hat{H}' = \omega_0 \delta \hat{b}\dagg\delta\hat{b} + (\omega-U/2)\delta\hat{a}\dagg\delta\hat{a} 
            +g\left(\delta\hat{b} + \delta\hat{b}\dagg\right)\left(\delta\hat{a}+\delta\hat{a}\dagg\right)\, . 
\end{eqnarray}
Comparing this with \eq{eq:H_HP}, we see that they agree upon inserting $\bar{j}_z=-1/2,\bar{\alpha}=0$.

\subsection{The super-radiant phase}

For the super-radiant phase, we have that $a_0,b_0$ behave like  $\mathcal{O}(\sqrt{N})$. We therefore expand $\hat{\xi}$ to order $\mathcal{O}(1/\sqrt{N})$, i.e.
\begin{eqnarray}
  \hat{\xi} \simeq \sqrt{k} - \half\frac{b_0(\delta\hat{b}+\delta\hat{b}\dagg)}{k} - \half\frac{\delta\hat{b}\dagg\delta\hat{b}}{\sqrt{k}} - \frac{\sqrt{k}}{8}\frac{b_0^2(\delta\hat{b}+\delta\hat{b}\dagg)^2}{k^2}\, ,
\end{eqnarray}
where $k \equiv N-b_0^2$. Neglecting constant terms, and terms that vanish as $N\to\infty$, we then arrive at the following effective Hamiltonian:
\begin{eqnarray}
  H' &= \left(\omega - \frac{U}{2} + \frac{U}{N}b_0^2\right)\delta\hat{a}\dagg\delta\hat{a} \nonumber \\
     &+ \left(\omega_0-\frac{gb_0(a_0+a_0\conj)}{\sqrt{Nk}}+\frac{U}{N}|a_0|^2\right)\delta\hat{b}\dagg\delta\hat{b} \nonumber \\
     &+2g\sqrt{\frac{k}{N}}b_0(\delta\hat{a}+\delta\hat{a}\dagg)+\left(\omega-\frac{U}{2}+\frac{U}{N}b_0^2\right)(a_0\conj\delta\hat{a}+a_0\delta\hat{a}\dagg)\nonumber \\
     &+\Big[2g\frac{a_0+a_0\conj}{\sqrt{Nk}}\left(\frac{N}{2}-b_0^2\right) + \omega_0b_0 + \frac{U}{N}b_0|a_0|^2\Big](\delta\hat{b}+\delta\hat{b}\dagg)\nonumber \\
     &+\frac{2g}{\sqrt{Nk}}\left(\frac{N}{2}-b_0^2\right)(\delta\hat{a}+\delta\hat{a}\dagg)(\delta\hat{b}+\delta\hat{b}\dagg)\nonumber \\
     &+\frac{U}{N}b_0(a_0\conj\delta\hat{a}+a_0\delta\hat{a}\dagg)(\delta\hat{b}+\delta\hat{b}\dagg)\nonumber \\
     &-\frac{gb_0(a_0+a_0\conj)}{4k^2}\sqrt{\frac{k}{N}}(b_0^2+2k)(\delta\hat{b}+\delta\hat{b}\dagg)^2\, .
\end{eqnarray}
By making use of (cf. \eq{eq:SR})
%
%
\begin{eqnarray}
  \frac{b_0}{\sqrt{N}} = \sqrt{\half+\bar{j}_z}\, ,
\end{eqnarray}
and
\begin{eqnarray}
  \frac{a_0}{\sqrt{N}} = -\frac{2g\sqrt{1/4-\bar{j}_z^2}}{\omega+U\bar{j}_z-i\kappa}\, ,
\end{eqnarray}
%
 one now finds the commutator
\begin{eqnarray}
  &\left[2g\sqrt{\frac{k}{N}}b_0(\delta\hat{a}+\delta\hat{a}\dagg)+\left(\omega-\frac{U}{2}+\frac{U}{N}b_0^2\right)(a_0\conj\delta\hat{a}+a_0\delta\hat{a}\dagg),\delta\hat{a}\right] \nonumber \\
  &= -i\kappa a_0\, ,
\end{eqnarray}
 and that
\begin{eqnarray}
  2g\frac{a_0+a_0\conj}{\sqrt{Nk}}\left(\frac{N}{2}-b_0^2\right) + \omega_0b_0 + \frac{U}{N}b_0|a_0|^2 = 0\, .
\end{eqnarray}
Hence, the terms in $H'$ that are linear in $\delta\hat{a}^{(\dagger)},\delta\hat{b}^{(\dagger)}$ will vanish in the Heisenberg equations of motion. Thus, after inserting the expressions for $a_0,b_0$ given above, we find that we can use Eqs.(\ref{eq:fluct_time_a})--(\ref{eq:fluct_time_b}) with the Hamiltonian as given in \eq{eq:H_HP}.

\section{\label{sect:app_fourier}Time-delayed feedback in linear quantum systems}

Many systems of interest in quantum optics have linear Heisenberg equations of motion, and they are typically treated by introducing Fourier transformed fields \cite{Walls08}. We extended this treatment to linear systems with feedback, as described by the following general form:
\begin{eqnarray}\label{eq:linearfb}
  \frac{d\hat{\mathbf{a}}(t)}{dt} = \mathbf{A} \cdot \hat{\mathbf{a}}(t) - \mathbf{\Gamma} \cdot \hat{\mathbf{a}}(t) + \sum_i \mathbf{K}_i \cdot \hat{\mathbf{a}}(t-\tau_i) - \sqrt{2\mathbf{\Gamma}} \cdot\hat{\mathbf{a}}_\text{in}(t)\, ,
\end{eqnarray}
where $\mathbf{a} = (\hat{a}_1,\hat{a}_1\dagg,\dots,\hat{a}_n,\hat{a}_n\dagg)$ is a vector of field modes and their adjoints, $\mathbf{A}$ is a matrix coupling the different fields, $\mathbf{\Gamma} = \text{diag}(\kappa_1,\kappa_1,\dots,\kappa_n,\kappa_n)$ is a diagonal matrix of decay rates, similarly $\sqrt{2\mathbf{\Gamma}} = \text{diag}(\sqrt{2\kappa_1},\sqrt{2\kappa_1},\dots,\sqrt{2\kappa_n},\sqrt{2\kappa_n})$ gives the coupling to the input fields $\hat{\mathbf{a}}_\text{in} = (\hat{a}_{1,\text{in}},\hat{a}_{1,\text{in}}\dagg,\dots,\hat{a}_{n,\text{in}},\hat{a}_{n,\text{in}}\dagg)$, and the matrix $\mathbf{K}_i$ couples the system to the feedback fields $\hat{\mathbf{a}}(t-\tau_i)$.

Next, we introduce the Fourier transforms
\begin{eqnarray}
  \tilde{O}(\nu) = \frac{1}{\sqrt{2\pi}} \int_{-\infty}^\infty \e^{i\nu t}\hat{O}(t) \dx{t}\, ,
\end{eqnarray}
and
\begin{eqnarray}
  \tilde{O}\dagg(-\nu) = \frac{1}{\sqrt{2\pi}} \int_{-\infty}^\infty \e^{i\nu t}\hat{O}(t)\dagg \dx{t}\, ,
\end{eqnarray}
%
%
for any operator $\hat{O}$. Since $1/\sqrt{2\pi}\int_{-\infty}^\infty \exp(i\nu t) a(t-\tau)\dx{t} = \exp(i\nu\tau)\tilde{a}(\nu)$, one finds the Fourier space equations of motion
\begin{eqnarray}
  \left[i\nu + \mathbf{A} -\mathbf{\Gamma} + \sum_i \mathbf{K}_i \e^{i\nu\tau_i}\right]\cdot\tilde{\mathbf{a}}(\nu) = \sqrt{2\mathbf{\Gamma}} \cdot\tilde{\mathbf{a}}_\text{in}(\nu)\, .
\end{eqnarray}
The system fields are thus solved in terms of the input fields by inverting the matrix on the left hand side.

We will now return to our system of study, i.e.  Eqs. (\ref{eq:fluct_time_a}) and (\ref{eq:fluct_time_b}), which read:
\begin{eqnarray}
  \frac{d\delta{\hat{a}}}{dt} &=& -i\omega_a\delta\hat{a} - i\big(\lambda_1-i\lambda_2\big)\big(\delta\hat{b}+\delta\hat{b}\dagg\big) -\kappa \delta\hat{a} \nonumber \\
                      &&+k\big(\delta\hat{a}(t-\tau)-\delta\hat{a}(t)\big) -\sqrt{2\kappa}\hat{a}_\text{in}(t)\, , 
\end{eqnarray}
\begin{eqnarray}
  \frac{d\delta{\hat{b}}}{dt} = -i\omega_b\delta\hat{b} - i\lambda_1\big(\delta\hat{a}+\delta\hat{a}\dagg\big) + \lambda_2\big(\delta\hat{a}-\delta\hat{a}\dagg\big) - i\frac{\chi}{2}\big(\delta\hat{b}+\delta\hat{b}\dagg\big)\, .
\end{eqnarray}
These equations are indeed of the form \eq{eq:linearfb}, and we can apply the results above. We find the following equations in Fourier space
%
%
\begin{eqnarray}
  \nu \delta\tilde{a}(\nu) =& \big(\omega_a -i\kappa\big)\delta\tilde{a}(\nu)
                            + \big(\lambda_1-i\lambda_2\big)\big(\delta\tilde{b}(\nu) + \delta\tilde{b}\dagg(-\nu)\big)\label{eq:fourier_fluct_a}\nonumber \\
                        &+ ik\big(e^{i\nu\tau}-1\big)\delta\tilde{a}(\nu) - i\sqrt{2\kappa} \tilde{a}_\text{in}(\nu)\, ,
\end{eqnarray}
and
\begin{eqnarray}
  \nu \delta\tilde{b}(\nu) = \omega_b \delta\tilde{b}(\nu) &+ \lambda_1 \big(\delta\tilde{a}(\nu) + \delta\tilde{a}\dagg(-\nu)\big)
       + i\lambda_2 \big(\delta\tilde{a}(\nu) - \delta\tilde{a}\dagg(-\nu)\big) \label{eq:fourier_fluct_b}\nonumber \\
                            &+ \frac{\chi}{2}\big(\delta\tilde{b}(\nu) + \delta\tilde{b}\dagg(-\nu)\big) \, .
\end{eqnarray}
The solution to these algebraic equations are
\begin{eqnarray}
  \fl
  \delta\tilde{a}(\nu) = \frac{i\sqrt{2\kappa}}{D(\nu)} \big({\left[2\omega_bG^2+N(\nu)\big(\nu^2-\omega_b(\omega_b+\chi))\right] \tilde{a}_\text{in}(\nu) + 2\omega_bG^2 \tilde{a}\dagg_\text{in}(-\nu)\big)}\, ,
\end{eqnarray}
and
\begin{eqnarray}
  \fl
  \delta\tilde{b}(\nu) = \frac{i\sqrt{2\kappa}}{D(\nu)} (\nu+\omega_b) \big({G\conj N(\nu)\tilde{a}_\text{in}(\nu) + G N\conj(-\nu)\tilde{a}\dagg_\text{in}(-\nu)\big)}\, ,
\end{eqnarray}
where
%
%
\begin{eqnarray}
  G = \lambda_1-i\lambda_2\, ,\\
  N(\nu) = \omega_a + i\left[\kappa - i\nu - k\big(\e^{i\nu\tau} -1\big)\right]\, ,
\end{eqnarray}
and
\begin{eqnarray}
  D(\nu) &= \big({\omega_a^2 + \left[\kappa -i\nu - k\left(\e^{i\nu\tau}-1\right)\right]^2\big)}\big(\nu^2-\omega_b(\omega_b+\chi)\big)\nonumber \\ &+ 4|G|^2 \omega_a \omega_b\, .
\end{eqnarray}
%
%
These solutions can then be used to compute the quantum fluctuation in steady state:
\begin{eqnarray}\label{eq:ada_ss}
  \braket{\delta \hat{a}\dagg \delta \hat{a}}_\text{ss} = \frac{1}{2\pi} \int_{-\infty}^\infty \int_{-\infty}^\infty \braket{\delta \tilde{a}\dagg(\nu)\delta \tilde{a}(\nu')} \dx{\nu}\dx{\nu'} \nonumber \\
                                                        = \frac{\kappa}{\pi} \int_{-\infty}^\infty 4\omega_b^2\left|\frac{G^2}{D(\nu)}\right|^2 \left[1+\frac{k}{\kappa}(1-\cos\nu\tau)\right]\dx{\nu}\, ,
\end{eqnarray}
where we have used the Fourier space version of \eq{eq:corr_ain}, which can be expressed as
\begin{eqnarray}
  \braket{\tilde{a}_\text{in}(\nu)\tilde{a}\dagg_\text{in}(\nu')}& = \left[1+\frac{k}{\kappa}(1-\cos\nu\tau)\right]\delta(\nu-\nu')\, ,
\end{eqnarray}
while all other correlation functions vanish.
\newline


\end{document}